\documentclass[a4paper]{spie}  

\usepackage{amsmath,amsfonts,amssymb}
\usepackage[colorlinks=true, allcolors=blue]{hyperref}

\usepackage[english]{babel}
\usepackage[utf8]{inputenc}
\usepackage[]{graphicx}
\usepackage{microtype}
\usepackage{epsfig}
\usepackage{float}
\usepackage{subfig}
\usepackage{gensymb}
\usepackage{todonotes}

\usepackage{natbib}
\bibpunct{(}{)}{;}{a}{}{,}

\graphicspath{{Pics/}}
\usepackage{txfonts}
\usepackage{multicol}
\usepackage{enumerate}
\usepackage{tikz}
\usetikzlibrary{arrows.meta, calc, chains, positioning, quotes, shapes}
\tikzset{
alias path picture bounding box/.code=%
    \pgfnodealias{#1}{path picture bounding box},
CNTRL/.style = 
{
          > = Triangle,
shorten <>/.style = {shorten <=##1, shorten >=##1},
block/.style = {rectangle, fill=##1,
    draw = blue, minimum height=8mm, minimum width=16mm,
    outer sep = 0mm},
       block/.default = white!100,
dot/.style = {fill=##1,
    circle, inner sep=0mm, outer sep=0mm, minimum size=1mm,
    node contents={}},
edot/.style = {fill=##1,
    circle, inner sep=0mm, outer sep=0mm, minimum size=0.mm,
    node contents={}},
mlt/.style = {fill=##1,
    rectangle, minimum size=6mm,
    path picture={%
    \tikzset{alias path picture bounding box=@}
    \draw[very thick,shorten <>=1.5mm,-]
    (@.north west) edge (@.south east)
    (@.south west)  --  (@.north east);
                },
            node contents={}},
sum/.style = {fill=##1,
    circle, minimum size=2mm,
    path picture={%
    \tikzset{alias path picture bounding box=@}
    \draw[very thick,shorten <>=1mm,-]
    (@.north) edge (@.south)
    (@.west)   --  (@.east);
                },
            node contents={}},
decision/.style = {diamond, fill=##1, draw = blue,
    minimum height=8mm, minimum width=12mm,
    outer sep = 0mm},
     decision/.default = white!100,
cloud/.style = {ellipse, fill=##1, draw = blue,
    minimum height=8mm, minimum width=16mm,
    outer sep = 0mm},
    cloud/.default = white!100,
realm/.style = {rounded rectangle, text=background, fill=##1,
    minimum height=8mm, minimum width=16mm,
    outer sep = 0mm},
}
}

\title{Fiber-coupling of Fourier Transform Spectrographs}

\author[a]{Sch\"afer, S.}
\author[b]{Huke, P.}
\author[a]{Meyer, D.}
\author[a]{Reiners, A.}

\affil[a]{Georg-August Universit\"at G\"ottingen, Institut f\"ur Astrophysik, Friedrich-Hund-Platz 1, 37077 G\"ottingen, Germany}
\affil[b]{University of applied sciences Emden/Leer, Institute for Laser and Optics, Constantia-Platz 4, 26723 Emden, Germany}
\authorinfo{Further author information: (Send correspondence to Sebastian Sch\"afer)\\ Sebastian Sch\"afer: E-mail: schaefer@astro.physik.uni-goettingen.de, Telephone: +49 (0)551 39 25068}

\begin{document} 
\maketitle

\begin{abstract}

Fourier Transform Spectrographs (FTS) are versatile tools for measuring accurate, high resolution spectra. They are internally calibrated by a reference laser that runs in parallel to the science light. Therefore it is crucial to properly align these two beams with respect to each other. We show how this can be achieved by feeding a part of the reference light into the optical path of the science beam. 

For astronomical applications it's often useful to use optical fibers. We present a coupling setup for our Bruker Optics IFS 125 FTS, consisting of (1) two hexagonal input fibers,  (2) dichroic beam-combining for measuring two light sources simultaneously and (3) optimized optics to match the original Bruker design. The hexagonal shape of the fiber cores secures sufficient mode scrambling inside the fibers, resulting in constant beam parameters and a more homogeneous illumination of the entrance aperture of the FTS. 
\end{abstract}


\section{Fourier Transform Spectrographs}
There are many good books and publications about Fourier Transform Spectrographs (FTS), see e.g. \cite{Davis2001} and \cite{Griffiths2007}. In the following section, the theory is outlined only as much as it is needed to explain how the recorded interferogram is affected by a fiber coupling. For an overview on our FTS setup see Fig.\,\ref{fig:Setup}
\subsection{Principle of Operation}
\label{ssec:PoP}
The FTS is a Michelson-interferometer and the electric field incident on the beamsplitter is a superposition of many electric fields $E_\mathrm{in,k}$ with amplitude $A_k$ and single frequency $\nu_k$: 
\begin{equation}
E_\mathrm{in}=\sum_{k=1}^NA_k\cdot e^{-i\nu_kt+\phi_k(t)}.
\label{eq:singEfield}
\end{equation}
The phase $\phi_k(t)$ of the respective \textit{frequency component k} consists of a constant offset, a periodically changing term and a term describing the noise.

\begin{figure}[h!]
  \centering
  \fbox{\includegraphics[width=1.\linewidth]{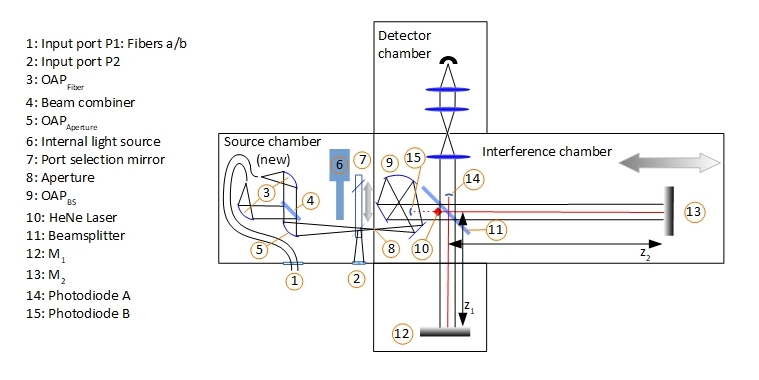}}
  \caption{Functional sketch of the setup. The FTS (a Bruker IFS125HR) consists of a source chamber, an interference chamber and a detector chamber. Two input ports (1 and 2) feed the "science light" into the FTS. Input 1 consists of two fibers (see Hex1 at tab.\,\ref{tab:Fibers}) that are collimated using a off-axis parabolic (OAP) mirrors (3) and combined with a dichroic mirror (4). Another OAP (5) is focussing the beam thorugh the aperture (8). Alternatively, input 2 can be used which directly accepts a focussed beam up to F/4.18. For more details on the source chamber see Sec.\,\ref{sec:coupling}.
  A flat folding mirror redirects the light towards the collimating off-axis parabolic mirror (9). The collimated light is divided with the beamsplitter (11), travels the paths $z_{1/2}$ to two retroreflector mirrors $M_{1/2}$ (12,13), recombined at the beamsplitter and the interference is recorded with the detector. The \textit{HeNe}-Laser (10) (customized SL04 from SIOS) is used as the reference laser inside the FTS (red optical path) with its own detectors: photodiodes A (14) and B (15).}
     \label{fig:Setup}%
\end{figure}

In the FTS the electric field is divided into two by the beamsplitter with reflectivity $r_k$ and transmission $t_k$. After propagation and retroreflection at the mirrors $M_{1,2}$, the electric fields $E_a(t_1),E_b(t_2)$ are recombined at the beamsplitter. The photodiode (PD), records the interference signal 
\begin{equation}
    I(\Delta t)=\langle E_a(t_1)+E_b(t_2) \rangle^2=
    \langle\sum_{k=1}^N A_{a,k}(t_1)\cdot e^{-i\left(\nu_k t_1+\phi_{k,a}(t_1)\right)}+
    \sum_{k=1}^N A_{b,k}(t_2)\cdot e^{-i\left(\nu_k t_2+\phi_{k,b}(t_2)\right)}\rangle^2,
\label{eq:int2field}
\end{equation}
with the two amplitudes $A_{a/b,k}=r_kt_k\cdot A_k$ for frequency component $\nu_k$. For simplification the end mirror reflectivities were omitted. Obviously, the efficiency of the instrument varies with $r/t_k$. The two electric fields contain different phases $\phi_{a,b}$ as a result of:
\begin{itemize}
\item[(a)] The optical path length difference $OPD = 2(z_2-z_1)$:

Typically, beamsplitters are only one side partial reflective while the other is antireflection coated. Due to this, one electric field (the second arm in Fig.\,\ref{fig:Setup} $E_b$) traverses the material three times and the other ($E_a$) only once. The frequency-dependent refractive index of the material adds phase to the electric fields and the paths become frequency dependent. In addition the frequency-dependent refractive index of the atmosphere in the FTS affect the beam path. 
\item[(b)] phase variations $\delta\phi_{\mathrm{in}}(t)$ of the incident electric field which happened during $\Delta t$ given by:
\begin{equation}
    \Delta t = t_2-t_1=2\frac{\Delta z}{c}=\frac{OPD}{c}.
\label{eq:FTSFilt}
\end{equation}
\end{itemize}
The result of Eq.\,\ref{eq:int2field} consists of two correlation terms:
\begin{equation}
    I(\Delta t) = \underbrace{\left(A_a\star A_b\right)(\Delta t)}_{Env_{\mathrm{Cor}}}\cdot\left(1+\underbrace{\cos\left(\Delta\phi(\Delta t)\right)}_{Car_{\mathrm{Cor}}}\right).
\label{eq:CrossCorInt}    
\end{equation}
At $z_2=z_1$ the recorded interferogram is an auto-correlation (see \cite{Diels2006}), elsewhere it is a cross-correlation (see \cite{Ye2002}).  For a light source emitting electric fields with constant or slowly varying amplitudes the term $Env_{\mathrm{Cor}}=I/2$ can be subtracted from the interferogram.

The remaining correlation term $Car_{\mathrm{Cor}}$ is a superposition of \textit{cosine}-terms with phase differences depending on the optical path difference. The power spectral density can be computed by using a Fourier transform and a phase-correction algorithm, to compensate for phase-noise and the frequency-dependent OPD (see \cite{Mertz1966} and \cite{ Learner1995}). Further, additive noise sources, like detector, digitalization or photon noise are well known and can be minimized \citep{Davis2001}.

\subsection{Apodization}
For an unperturbed, continuous electric field with single frequency component $\nu_k$ the finite length of the interferometer limits the measurement. Truncation is an apodization with a boxcar-function. Its Fourier representation is a sinc-function \citep{Davis2001} which is the instrumental line shape (ILS) \citep{Hase1999}. To reduce the ringing of the ILS other apodization-function are better suited, \cite{Naylor2007} give a nice overview on apodization-functions.\\
The electric fields $E_{in}(t_1/t_2)$ coupled into the FTS are jointly stationary in a wide sense \citep{Goodman2000} within the coherence time $\tau_{\mathrm{coh}}$ of the source, when the stochastic processes of light generation and perturbation on the light path are Gaussian. The phase variations of the electric field acts on the dc-filtered interferogram $Car_{\mathrm{Cor}}$ and it can be shown that the measured intensity is: 
\begin{equation}
    I_{\mathrm{AC}}(\Delta t) = \underbrace{\frac{1}{\sqrt{2\pi\sigma^2}}e^{-\frac{\Delta t^2}{2\sigma^2}}}_{RP_{\mathrm{phase}}}\cdot\underbrace{\cos\left(\Delta\phi(\Delta t)\right)}_{Car_{\mathrm{Cor}}}.
\label{eq:GaussInt}    
\end{equation}
As a result, the contrast:
\begin{equation}
    C=\frac{I_\mathrm{AC,max}-I_\mathrm{AC,min}}{I_\mathrm{AC,max}+I_\mathrm{AC,min}}
    \label{eq:Contrast}
\end{equation}
of the interferogram decreases due to the phase variations. The stochastic processes work effectively like an apodization.

\subsection{Fiber-coupling}
\label{ssec:fiber-FTS}
The finite diameter and divergence angle of an extended light source affects the interferogram \citep{Saarinen1992}. To prepare the light with well-defined parameters, the light is filtered with an aperture (see item (8) in Fig.\,\ref{fig:Setup}) with diameter $d$. The divergence angle is determined by the focal length (and diameter) of the collimator in front of the beamsplitter (see Fig.\,\ref{fig:Setup}, item (9)). Stationary deformations with low spatial frequencies, i.e. aberrations, affect the instrumental line shape, see e.g. Ahro et. al. \cite{Ahro2000} for a nice overview. Higher spatial frequencies, e.g. speckles, deteriorate the interferogram, as their coherence volume can be rather small \cite{Dainty1975, Leendertz1970}.

The wavefront leaving a multimode fiber is a superposition of the fields propagating in the modes of the fibers. The contrast of the pattern is governed by the temporal behaviour and the coherence of the incident electric field. A high coherence creates a high contrast \citep{Berdague1982} and by observation of the fiber exit a spectrometer can be realized \citep{Coluccelli2016}. The residual wavefront is structured with high spatial frequencies and varies quickly. 

To smoothen the wavefront either very few or as many as possible modes should be addressed. This can be achieved with a broadband light sources with a low degree of coherence or, if the light source can not be chosen freely, by using fibers which distribute the energy between the modes more evenly, like hexagonal or octagonal fibers. These fibers work like beam-homogenizers. The residual changes of the wavefront may be filtered by the recording process through averaging.\\ 
To measure the effect on the interferogram, a part of the light of the highly coherent reference laser (see item (10) in Fig.\,\ref{fig:Setup}) was coupled into different fibers, see Table\,\ref{tab:Fibers}.

\begin{table}[h!]
\caption{Properties of fibers used with the FTS.} 
\label{tab:Fibers}
\begin{center}       
\begin{tabular}{|l|l|l|l|l|l|l|} 
\hline
\rule[-1ex]{0pt}{3ex} Name & \O \vspace{1pt} [$\mu m$] & NA & length [m] & Core-Shape & Material & Vendor\\
\hline
\rule[-1ex]{0pt}{3.5ex}  Hex1 & 525 & 0.26 & 1 & hexagonal & Fused silica & CeramOptec\\
\hline
\rule[-1ex]{0pt}{3.5ex}  Hex2 & 525 & 0.26 & 2 & hexagonal & Fused silica & CeramOptec\\
\hline
\rule[-1ex]{0pt}{3.5ex}  R1 & 200 & 0.22 & 1 & round & Fused silica, low OH & Thorlabs\\
\hline
\rule[-1ex]{0pt}{3.5ex}  R2 & 365 & 0.22 & 10 & round & Fused silica, low OH & Thorlabs\\
\hline
\rule[-1ex]{0pt}{3.5ex}  R3 & 600 & 0.39 & 2 & round & Fused silica, low OH & Thorlabs\\
\hline
\rule[-1ex]{0pt}{3.5ex}  PM1 & 4.5 & 0.22 & 1 & panda & Fused silica, low OH & Thorlabs\\
\hline
\end{tabular}
\end{center}
\end{table} 
In our setup (see Fig.\,\ref{fig:Setup}) the fibers were directly coupled to one of the HEX1-fiber at $P_1$, and the dichroic replaced with a mirror. As the light of the laser is travelling on the reference as well as on the science path, the impact from laser-specific effects, e.g. intensity variations, can be reduced.

\begin{figure}[h!]
\centering
\begin{minipage}{0.494\linewidth}
	\raggedleft
	\includegraphics[width=1\linewidth]{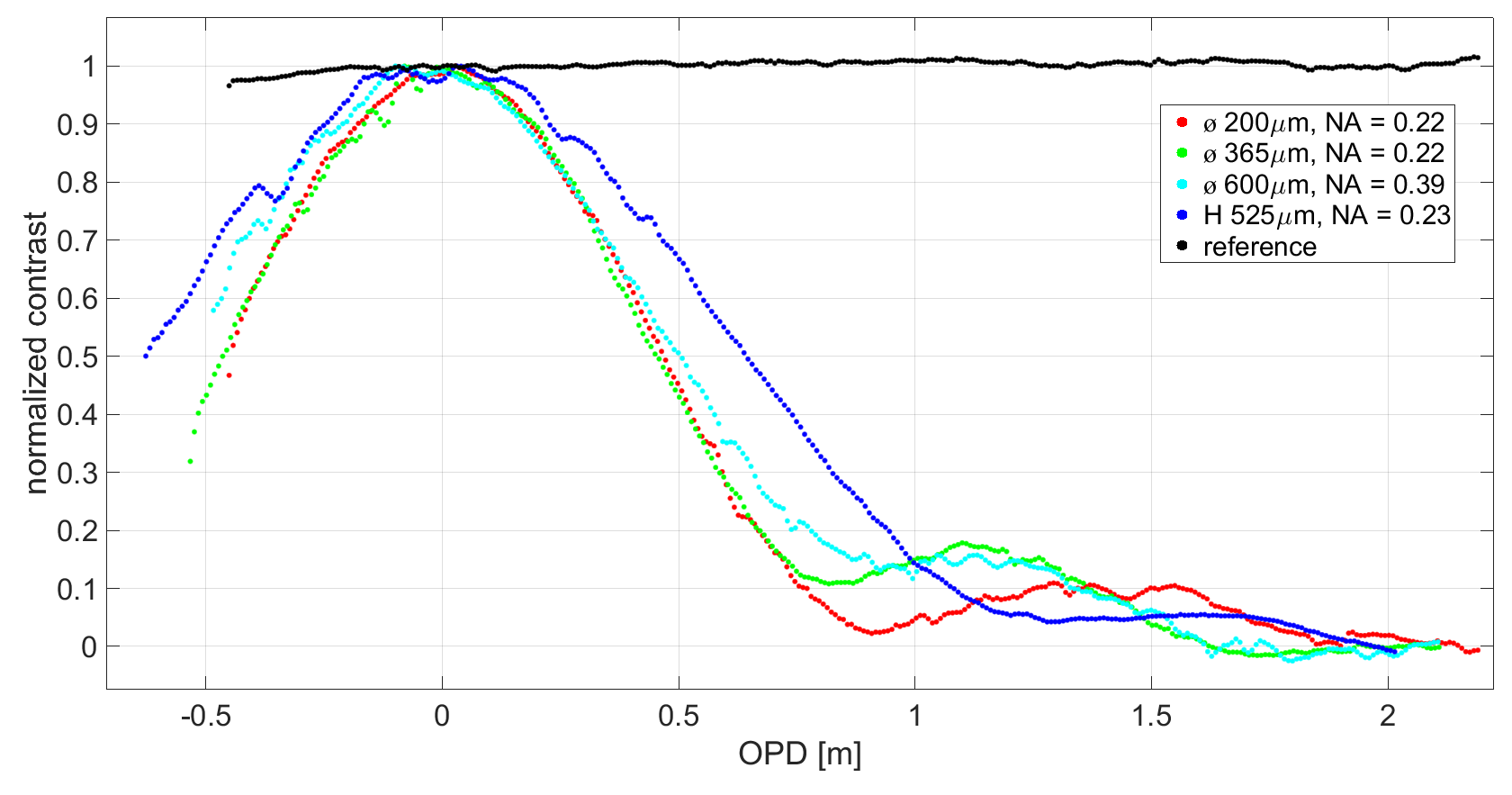}
\end{minipage}
\begin{minipage}{0.494\linewidth}
	\raggedright
	\includegraphics[width=1\linewidth]{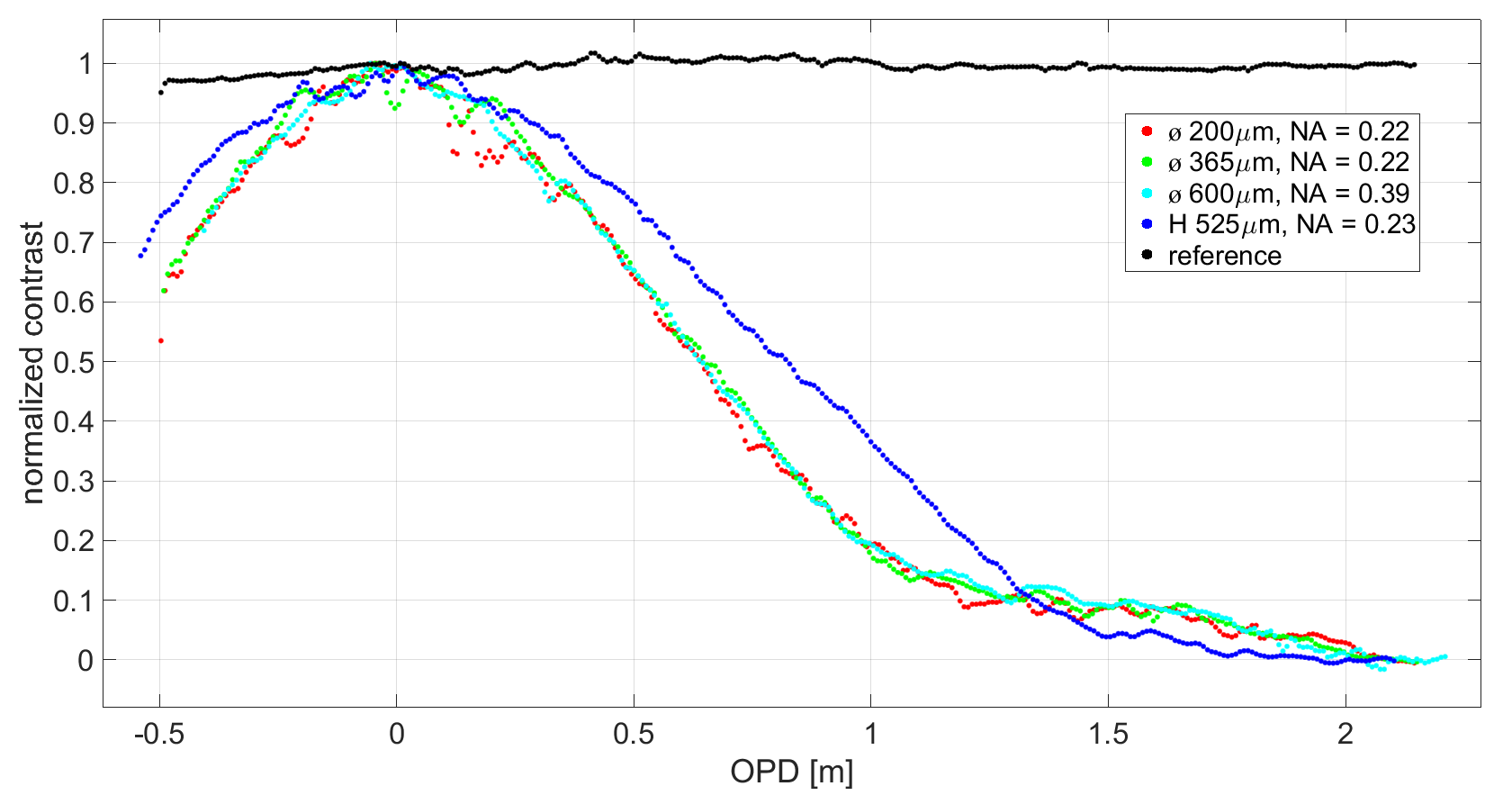}
\end{minipage}
\caption{Light of the reference laser is fed into the FTS on the Science path via the port P1 (left) and P2 (right), see figure \ref{fig:Setup}. The contrast $C$ of the recorded interferogram is computed using equation \ref{eq:Contrast} and normalized to the $OPD=0$-point.}
	\label{fig:PContrast}
\end{figure}

The laser has a linewidth of $\delta \nu \leq 1 $ MHz, which results in a coherence length of $L_{\mathrm{coh}}\leq100$ m. After this distance the contrast of the interferogram, (see Eq.\,\ref{eq:GaussInt}), drops to $I_{\mathrm{AC}}=1/2$. For the short measurement in the given interferometer the change in contrast should be negligible small which can be seen in the interferogram of the reference path. However, the science signal is affected by the fibers, see Fig.\,\ref{fig:PContrast}.

The contrast decreases as the incoupled light is perturbed by the $RP_{\mathrm{phase}}$-factor in Eq.\,\ref{eq:GaussInt}. As can be seen, all fibers affect the interferogram. In addition, the F2F-coupling into the Hex1-fiber decrease the contrast even further. The contrast of the SM-fiber is noisy, as the coupling into the fiber was not optimal. The fibers with circular-shaped core show a higher drop of the contrast compared to the fibers with a hexagonal core. This can be explained by the amount of modes available in the cores as well as the homogeneitiy of the beam, which is much better for the hexagonal-shaped cores. 
\subsection{Resolving power of the fiber-coupled FTS}
\label{ssec:ResPow}
\begin{figure}[h!]
  \centering
  \fbox{\includegraphics[width=0.85\linewidth]{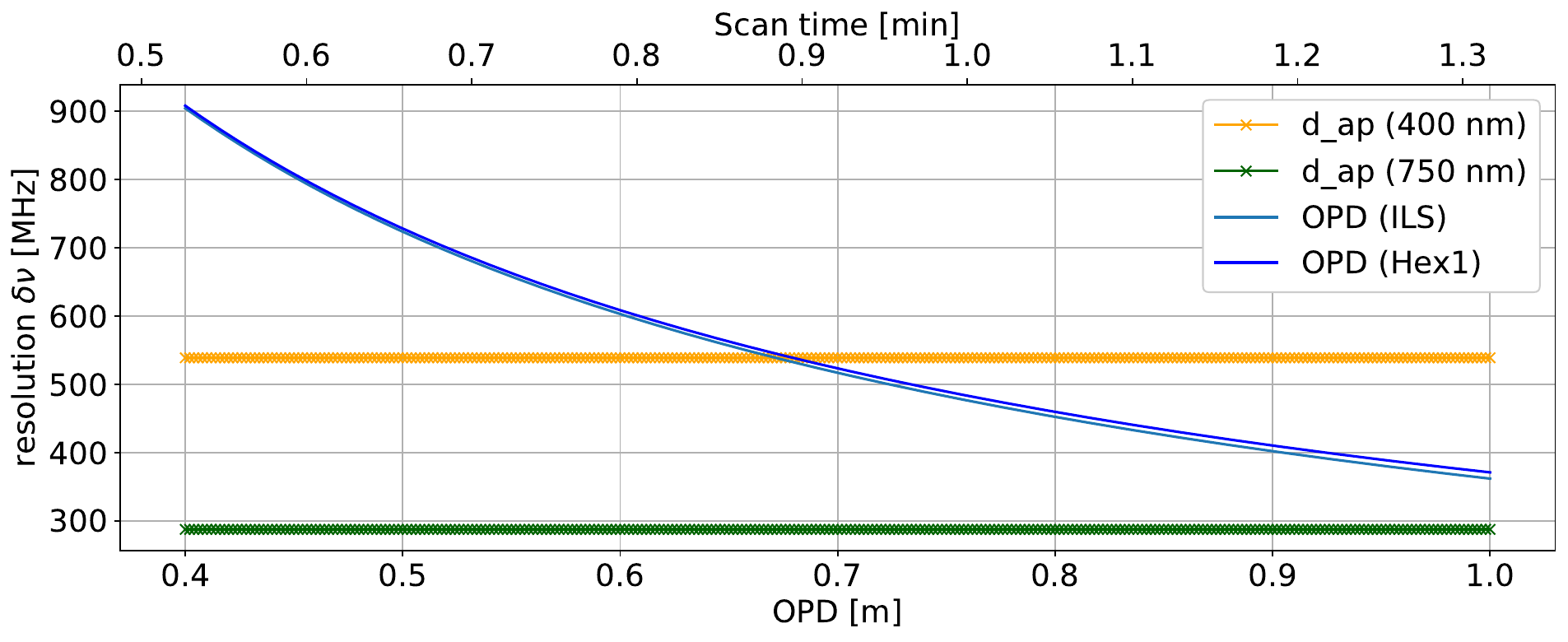}}
  \caption{Resolution of the FTS. The resolution is either limited by the OPD (ILS or Hex1 in case of fiber coupling) or by the aperture ($d_{\mathrm{ap}}=1$ mm) inside the FTS. The resolution was calculated for the lower and upper wavelength limit of the visible range. The scan time is calculated for a $v_{\mathrm{scan}}=20 $ kHz single-sided interferogram.}
 \label{fig:FTS_resolution}%
\end{figure}

The resolving power $R$ of an FTS at frequency $\nu$,
\begin{equation}
R\equiv\frac{\nu}{\delta\nu},
\label{eq:ResPrinc}
\end{equation}
is defined on the observable width $\delta\nu$ of an unresolved spectral line at frequency $\nu$. A line is unresolved if its line-width is significantly smaller than the width of the instrumental line shape (ILS). As resolution the full width at half maximum of the ILS $\delta\nu$ is taken. It can be limited by:
\begin{enumerate}
\item[(a)] Beam quality of the extended light source:

With the aperture ($d=1$ mm) and the focal length ($f_{\mathrm{Col}}=418$ mm of the collimating mirror inside the FTS, the resolving power can be computed (see \cite{Davis2001}) by:
\begin{equation}
R_{\mathrm{Col}}=8\frac{f^2}{d^2}=1.397.792
\end{equation} 
\item[(b)] Truncation:

An interferogram is always truncated at a certain $OPD$. This defines the sampling frequency $\nu_{\mathrm{sampling}}$ to 
\begin{equation}
\nu_{\mathrm{sampling}}=\frac{c}{OPD}.
\label{eq:OPDLim}
\end{equation}
The truncation works as a boxcar filter applied to the interferogram. The Fourier representation of a boxcar is a sinc-function. The full width at half maximum $sinc(u)=1/2$ defines the factor $\eta\equiv2u/\pi=1.20671$ \citep{Davis2001}. The frequency-independent resolution is given by
\begin{equation}
\delta\nu=\eta\cdot\nu_{\mathrm{sampling}}.
\end{equation}
With Eq.\,\ref{eq:ResPrinc} the resolving power for different frequencies can be calculated. 
\item[(c)] Apodization:

Apodization is often used to reduce the ringing of the sinc-function. Unfortunately, it also broadens the ILS and reduces the resolution. In the case of fibers, the effect of the fibers can be regarded as an apodization.
\end{enumerate}
In Fig.\,\ref{fig:FTS_resolution} the resolution is calculated for different $OPD$. The aperture ($d_{\mathrm{ap}}=1$ mm) is chosen to be slightly smaller than 2\,x\,\O \vspace{1pt} of the Hex1-fiber to allow for a maximum throughput (see Sec.\,\ref{ssec:efficiency}). For $OPD>0.68$ m the aperture becomes the limiting factor for the lower wavelengths. The apodization caused by the use of the Hex1-fiber leads to a small broadening of the ILS.

\newpage
\section{FTS Input Options}
\label{sec:coupling}

\subsection{Original IFS 125 FTS Source Chamber}
\label{ssec:original_setup}
The original source chamber (see Fig.\,\ref{fig:Setup} of the FTS) had two input ports: Input channel 1 is accepting parallel light with a beam diameter of up to 2\,inch (50.8\,mm). Behind the vacuum window the beam is focussed by the off-axis parabolic (OAP) mirror OAP1 (5), an 90\degree off-axis parabolic mirror (OAP) with D=2\,inch and f=250\,mm onto the pinhole (8) leading to the interferometer chamber. A flat folding mirror redirects the beam onto the final collimating 30\degree OAP with D=100\,mm and f=418\,mm (9). The now collimated beam is send via another flat folding mirror towards the beamsplitter.

Input channel 2 is accepting a F/4.18 beam (driven by OAP (9) with f=418\,mm, D=100\,mm ) . A motorized folding mirror can be moved, blocking the beam path of P1 and the input beam is redirected onto the pinhole. Afterwards the light follows the same path as for P1.

Additionally the FTS hosts an internal Halogen light source (6). Using another movable mirror (not shown) its light can be directed towards the aperture. In our case this light source is mainly used for testing, Bruker put it in to allow for transmission spectroscopy with an additional testing chamber attached.

For fiber-coupled light sources (3) the P1 input is much easier to setup: Using a standard fiber collimating OAP (4) by Thorlabs with D=22\,mm and an effective focal length (EFL) of 50.8\,mm (RC12SMA-P01) the light from the fiber is send through the vacuum window and onto the first internal OAP (5) that is focussing the light into the aperture (8). A flat folding mirror redirects the light onto a 2nd internal OAP (9) which is collimating the light and another flat folding mirror is sending it towards the main beamsplitter.

The main drawback of this setup is the magnification M of the fiber due to the focal lengths of OAP (5) and RC12SMA-P01: 
\begin{equation} 
M=\frac{f_{OAP5}}{f_{RC12SMA-P01}}=4.9 
\label{eq:Magnification}
\end{equation}
Since we are using a $525\,\mu m$ hexagonal fiber as input for most of our experiments, the spot on the pinhole gets magnified to about 2.5\,mm. Because we want to use a pinhole diameter of 1\,mm to achieve our desired resolving power (see Sect.\,\ref{ssec:ResPow}) this leads to a loss of about 84\% of the light at the pinhole.

\subsection{New Source Chamber Design Considerations}
\label{ssec:requriements}
For the new design a number of requirements have been identified:
\begin{itemize}
    \item Dual-Channel capability
    \item Increased efficiency
    \item Input selection stage (external)
\end{itemize}

\subsubsection{Dual-Channel capability}
The FTS reference HeNe laser is a customized SL04 from SIOS Messtechnik\footnote{specsheet at https://sios-de.com/products/stabilized-hene-lasers/stabilized-hene-laser-sl-04/} and employs dual longitudinal mode stabilization.

Therefore, a simultaneous drift measurement is needed in order to achieve sub-m/s precision. In contrast to high precision Echelle spectrographs, the FTS can't handle two separate inputs and simultaneously measure spectra for them. Therefore we opt for a dichroic beamsplitter setup in order to combine the \textit{science} and the \textit{calibration} signals. Both signals have their own respective input fiber and are mixed by a dichroic mirror inside the source chamber (see Sect.\,\ref{ssec:optic_layout}).

\subsubsection{Increased efficiency}
\label{ssec:efficiency}
Using a $525\,\mu m$ hexagonal fiber the original FTS source chamber setup has an efficiency of only 16\% (see Sect. \ref{ssec:original_setup}). This is undesirable for two reasons: 
\begin{itemize}
    \item In order to achieve a good signal-to-noise ratio (SNR) the countrate on the detector should always be close to 32,000. Therefore the signal has to be amplified. This can introduce additional noise sources and while it is still beneficial overall to use the amplification, minimizing the need for it would be even better. 
    \item One of the core uses of our FTS is to test and evaluate calibration sources for astronomical spectrographs. Since these spectrographs are quite efficient and exposure times can be up to several tens of minutes, their calibrations sources typically are quite faint. Thus wasting 84\% of the light inside the source chamber can cause significant problems in getting good SNR in the FTS spectra. Even with full amplification some sources produce only a fraction of the desired 32,000 counts on the detector.
\end{itemize}

Improving the efficiency means to reduce the diameter of the re-imaged fiber on the pinhole. The maximum pinhole diameter to achieve a resolution of 0.018 (approx. R=925,000 at 600\,nm) is 1\,mm. Since the fiber diameter is $525\,\mu m$ the magnification of the imaged  fiber should be about M=2. As shown in Eq.\,\ref{eq:Magnification}, the magnification is the ratio of the two focal lengths of the two OAPs used to first collimate and then focus the beam from the fiber. Additional constraints originate from the existing optics in the interferometer chamber (see Sect.\,\ref{ssec:original_setup}), available design volume (see Fig.\,\ref{fig:Source_chamber_new}) and available standard optics and optomechanics (Sec.\,\ref{ssec:optic_layout}).

\subsubsection{Input selection stage (external)}
The basic setup to select different calibration and science sources is to simply use a FC/PC-FC/PC mating sleeve from Thorlabs (ADAFC1). The FC/PC connection provides acceptable mechanical stability and repeatability. However, a much more sophisticated setup would be the usage of a light distribution point, similar to the ones used in the Expresso spectrograph and the one planned for the ELT-HIRES spectrograph. We will design and build a comparable small scale version as a prototype for the HIRES light distribution point \citep{Huke2017} in the near future. It shall be able to select between at least four different reference sources (Laser Frequency Comb \citep{Charsley2017}, Fabry-Pérot Etalon \citep{Wildi2010}, an Iodine cell \citep{Hatzes2010} and different HCLs \citep{Sarmiento2014, Redman2011} and five different science sources (e.g. our resolved and integrated Sun experiments, see\,\cite{Schaefer2020b}).

\subsection{New Optical layout}
\label{ssec:optic_layout}
The main limitations on how the new setup can be designed to fulfill the requirements are the design volume restrictions inside the source-chamber and the existing Bruker optics inside the interferrometer-chamber. Additionally, we limit the maximum optics diameter to 2\,inch to keep the overall costs low.

The layout of the original source chamber not only includes an OAP to focus the light entering the instrument through the parallel port P1, but also a setup of internal flatfield lamps and a motorized movable mirror to switch from P1 to P2 (the F/4.18 entrance window port). Therefore a significant part of the source chamber close to the pinhole is not available  for the new design. Essentially, nothing can be placed within a distance of 95\,mm of the pinhole, and horizontal access to the optical axis of the pinhole is very limited up to a distance of 160\,mm from the pinhole. 

There are only two optical elements to consider: thesize of the pinhole and the diameter and focal lenght of the OAP inside the interferrometer-chamber. The pinhole is mainly an issue with respect to the efficiency, so the image of the fibers on the pinhole should be close to 1\,mm (see Sect.\, \ref{ssec:efficiency}).\\
The OAP in the interferrometer-chamber has a diameter of d=100\,mm and a focal length f=418\,mm. Thus, the focussing OAP inside the source-chamber should be F/4.18 or slower. 

\begin{figure}[H]
  \centering
  \includegraphics[width=0.75\linewidth]{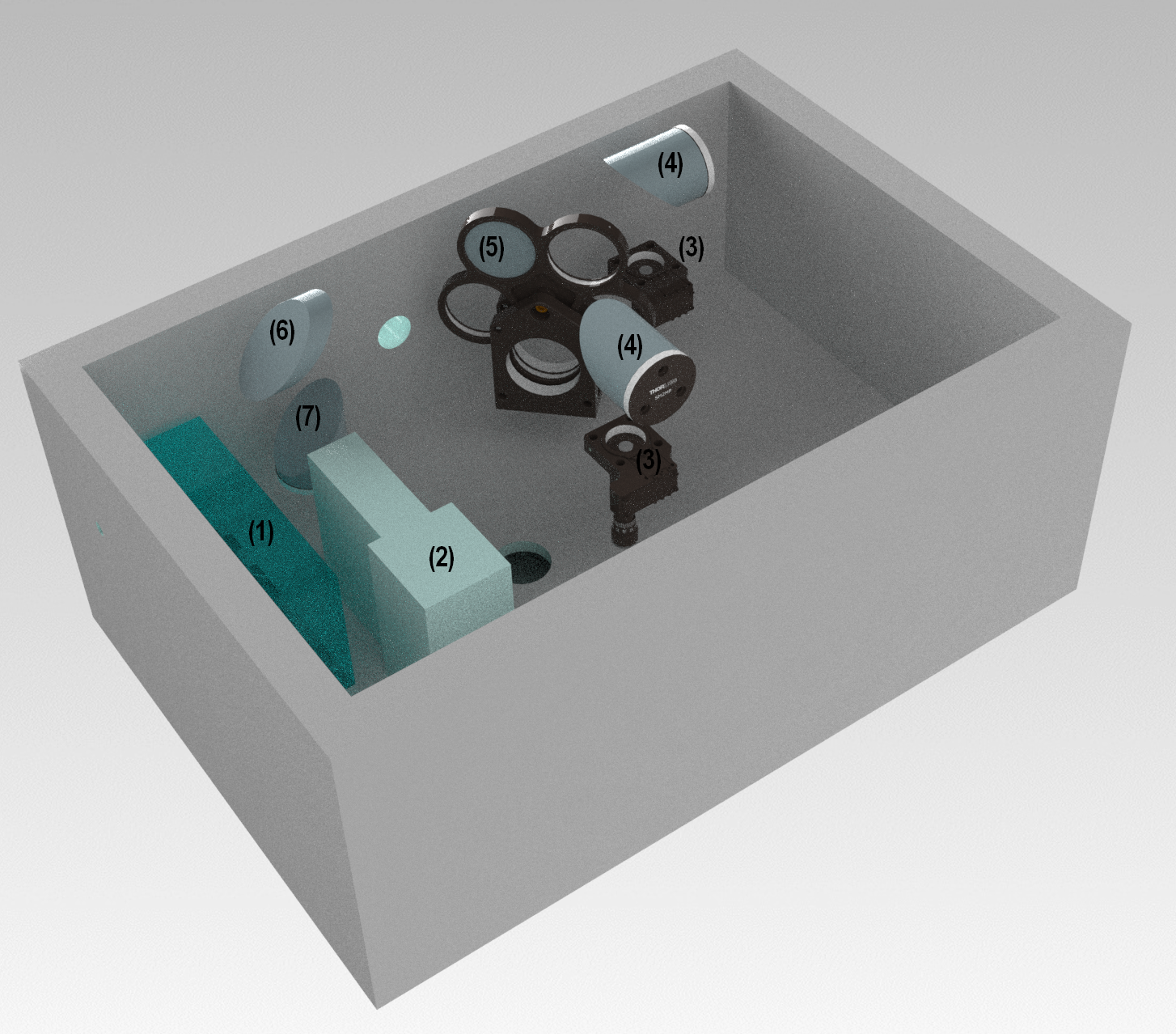}
  \caption{CAD-modell of the source chamber: (1) Space blocked by moving mirror; (2) Halogen lightsource housing; (3) Fiber ports; (4) OAP to collimate light from fibers; (5) Dichroic mirror inside filterwheel; (6) Folding mirror; (7) OAP focussing on the pinhole.}
 \label{fig:Source_chamber_CAD}%
\end{figure}

\begin{figure}[H]
  \centering
  \fbox{\includegraphics[width=0.75\linewidth]{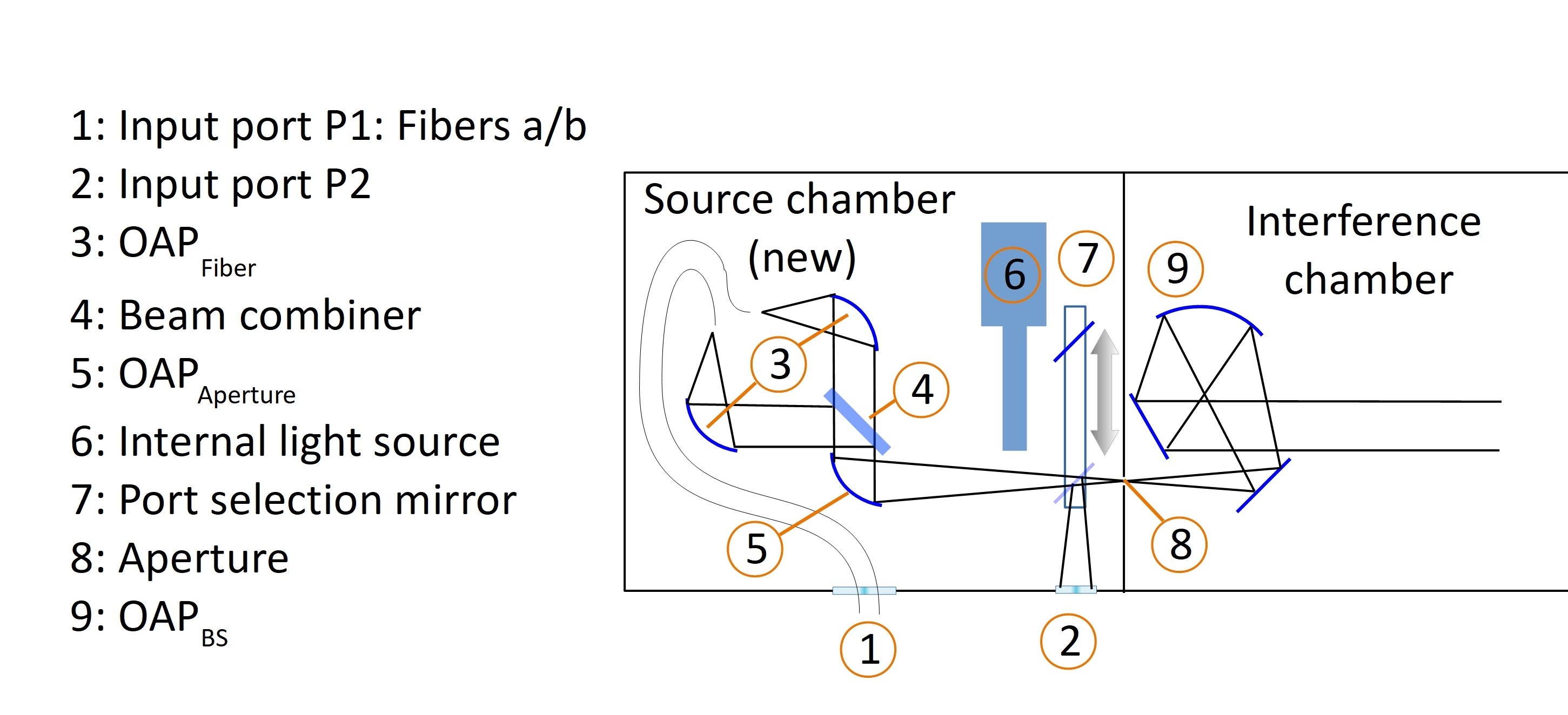}}
  \caption{Zoom into the new source chamber layout: The entrance window of input port 1 has been replaced by a fiber feedthrough hosting two fibers. Each fiber has its own OAP (3) to collimate the light. A dichroic beam combiner (4) is used to combine both channels as a function of wavelength, usually with a cut-off / cut-on wavelength of 800\,nm. The dichroic sits inside a filter wheel, different dichroics, a mirror (to only get light from one fiber) and an open position (to get only the light from the other fiber) can also be chosen depending on the experiment. OAP5 focusses the light into the aperture. Input channel two and the optics in the interference chamber stay unchanged.}
 \label{fig:Source_chamber_new}%
\end{figure}

\begin{figure}[h]
  \centering
  \includegraphics[width=0.85\linewidth]{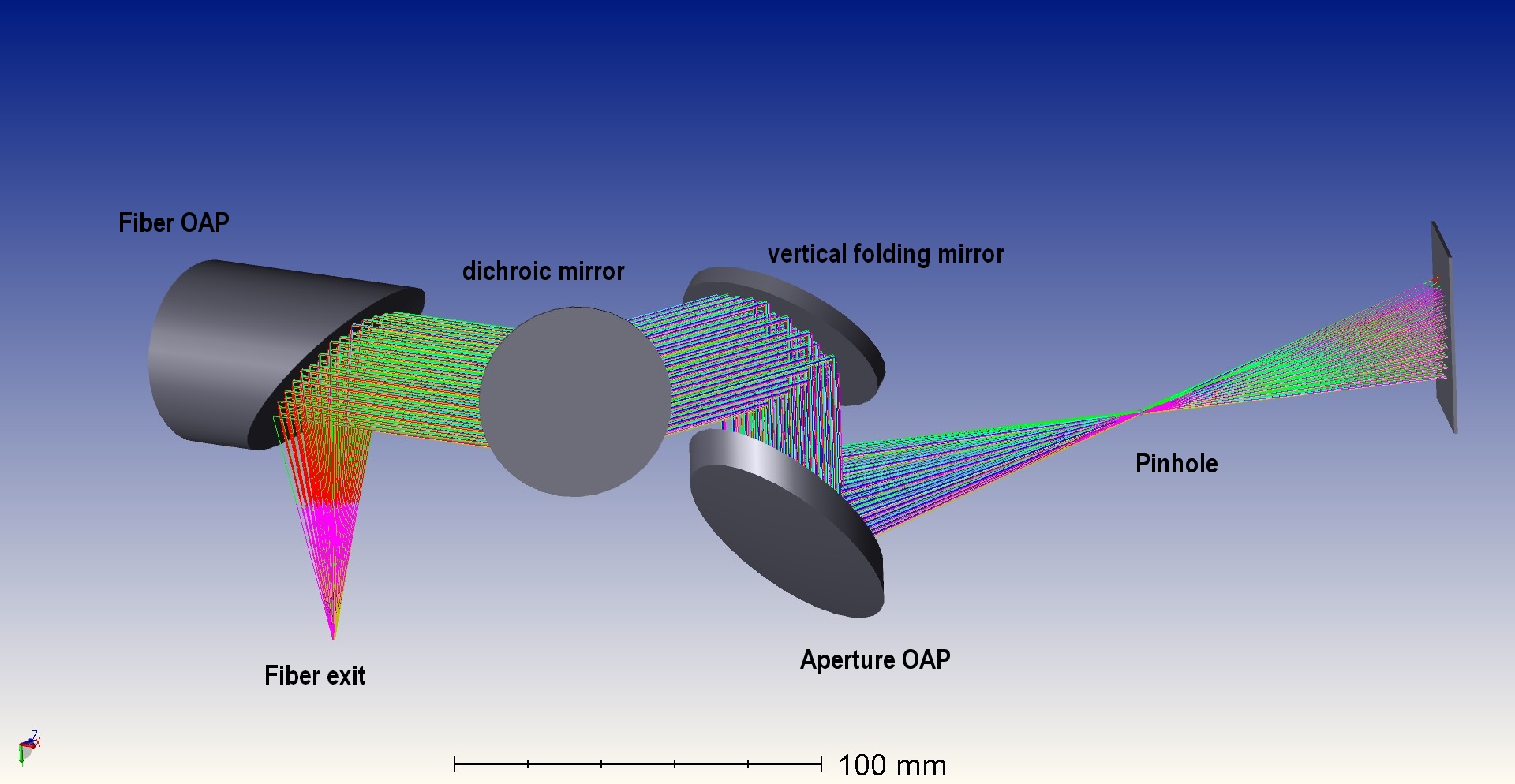}
  \caption{Zemax simulation of the new source chamber. Only one fiber exit and its corresponding OAP is shown. The light is collimated by the fiber OAP and reflected by the dirchroic (simulated as a simple 45\degree folding mirror). The 2nd fiber and its OAP (both not shown) are placed at a 90\degree angle with its beam hitting the dichroic from the backside. A vertical folding mirror is redirecting the beam onto the aperture OAP which is reimaging the fibers onto the pinhole. Behind the aperture the first mirror of the interference chamber is shown.}
 \label{fig:zemax}%
\end{figure}

Because we want to use a 2\,inch dichroic beamsplitter at 45\degree to combine two different light sources (see Sec.\,\ref{ssec:requriements}) our maximum beam diameter can only be $d=\frac{50.8mmm}{\sqrt{2}}=36\,mm$. Thus the focal length of the focussing OAP can be 150\,mm at max to still hit the interferrometer-chamber OAP with F/4.18. Consequently  we need to put it in the very limited space between the flatfield lamps and the wall and add an additional 45\degree fold mirror. Figure\,\ref{fig:Source_chamber_CAD} shows a CAD-modell of the source chamber, Fig.\,\ref{fig:zemax} contains a Zemax-simulation and Fig.\,\ref{fig:Source_chamber_new} is a sketch to illustrate the new setup.

We removed the parallel port vacuum window and installed a flange. The two hexagonal fibers ($d=525\mu m$, NA=0.26), are fed through the flange via feedthroughs (not shown) custom made by Ceramoptics. Both fibers have their own OAP with f=3\,inch (Edmund Optics OAPM 90 DEG 50.8 X 76.2 PROT SILVER 100A) to collimate the light. Thus the collimated beam is about 36\,mm in diameter. The two fiber exits and the optical axes of their OAPs are placed at a 90$^o$ angle to each other, meeting at the filter wheel.

A filter wheel (Thorlabs LCFW5) is used to mount different dichroic beamsplitters and a silver coated mirror. If necessary, one of the longpass beamsplitters (Thorlabs DMLP650L or DMLP805L) can be selected by the filter wheel to combine both beams. Alternatively, a mirror (Thorlabs PF20-03-P01) can be used to only get light from fiber 1 or an empty position can be used to only get light from fiber 2. Afterwards the beam is redirected by a 45\degree mirror (Thorlabs PFE20-P01) towards the focussing OAP which sits below the folding mirror. The OAP is a Thorlabs MPD269-P01, with f=6\,inch. Thus the setup images the fibers onto the pinhole with a magnification of M=2 and also fully illuminates the interferrometer chamber OAP with F/4.2.

\section{alignment procedure}
The end mirrors of the FTS are retroreflectors. Due to the fact that only one mirror, the stationary mirror $M_1$, see Fig.\ref{fig:Setup}, can be adjusted, the optical axis of the FTS is more or less fixed. The beam can not be tilted with respect to the optical axis of the moving retroreflected mirror without inducing signal deterioration. In principle the beam could propagate parallel to the optical axis defined by the center of the retroreflectors. In practice this also causes signal deterioration as both retroreflectors are not perfectly identical and flat mirrors. The question is now how to adjust a beam on the optical axis. The reference laser is using the same optical axis but is displaced to the edge of the retroreflectors. Thus it can be guided into the photodiodes A and B. The amplitude of the sinusodial signal of A and B must always be above a threshold, irrespective of the position of the moving mirror.

The alignment of a FTS is typically done by Bruker with a method based on white light fringes. In a first step the optical path of the reference laser is aligned with the available degrees of freedom before the beam splitter. This adjustement should be without moving the end mirror $M_{1}$. Therefore the signal of the photodiodes A and B at the beginning, the end and at $OPD=0$ (ZPD) is maximized. Due to the high degree of coherence of the reference laser, the signal should be constant.

In the next step a coherent light source is fed onto the science path by properly focusing onto the aperture at the entrance of the FTS. Therefore the OAP$_{Aperture}$ in the source chamber (see Fig.\ref{fig:Source_chamber_new})needs to be aligned so that the propagation direction of the beam with respect to the aperture plane is perfecty perpendicular. The three mirrors behind the aperture can be used to adjust the beam onto the optical axis.  The interference pattern at the output of the beamsplitter should look like ripples on the surface of a pond (sometimes referred to as Haydinger fringes), when $OPD \neq0$. During scanning the number of rings increases with OPD, but the pattern remains undisturbed otherwise. Next a continuous light source, e.g. a halogen lamp, is fed onto the optical path. The process of alignment is iterative: Alignment of a degree of freedom of the FTS, measurement of the phase spectrum and analysis of the instrumental line shape (ILS). If the ILS is distorted further alignment takes place. Typically, the fixed end mirror $M_2$ has to be aligned as well. As a consequence the reference laser needs to be adjusted again.  After some iterations the phase spectrum should be smooth. In the power spectrum the line profiles of e.g. oxygen lines in the atmosphere should be symmetric. The most important steps of the procedure is shown in the left flow chart in Fig.\,\ref{fig:bruker_align}.  

\begin{minipage}{0.5\textwidth}
\begin{tikzpicture}[CNTRL,
    node distance = 5mm and 15mm,
      start chain = A going below, every node/.style={scale=0.8}
                        ]
\node (block_1) [block, on chain=A]{\begin{tabular}{@{}cc}
     &Align the reference laser\\
     &using input degrees of freedom.
\end{tabular}};
\node (ed_1) [edot, left = of block_1];
\node (dec_1) [cloud, below = of block_1]{\begin{tabular}{@{}cc} &Amplitude A and B\\& above threshold\end{tabular}};
\node (block_2) [block, below = of ed_1]{Align $M_1$};
\node (block_3) [block,below = of dec_1]{Input aux. laser};
\node (block_4) [block,below = of block_3]{\begin{tabular}{@{}cc}Observe "Haydinger fringes" \\& \end{tabular}};
\node (dec_3) [cloud, below = of block_4]{Fringes concentric};
\node (block_4a) [block, left = of dec_3]{Mirror alignment};
\node (block_5) [block,below = of dec_3]{Input continuous light source};
\node (block_6a) [block,below = of block_5]{Measure interferogram};
\node (block_6) [block,below = of block_6a]{Analyze ILS};
\node (dec_2) [cloud, below = of block_6]{ILS};
\node (block_7) [block, left = of dec_2]{Mirror alignment};
\node (block_8) [block, below = of dec_2]{FTS aligned};
\draw[->] (block_1) edge  (dec_1);
\draw[->] (dec_1) -|  node[pos=0.3, above]{No}(block_2);
\draw[-] (block_2) edge (ed_1);
\draw[-] (ed_1) edge (block_1);
\draw[->] (dec_1) edge  ["Yes"](block_3);
\draw[->] (block_3) edge  (block_4);
\draw[->] (block_4) edge  (dec_3);
\draw[->] (dec_3) edge ["True"] (block_5);
\draw[->] (dec_3) edge ["False"](block_4a);
\draw[->] (block_4a) |- (block_4);
\draw[->] (block_5) edge  (block_6a);
\draw[->] (block_6a) edge  (block_6);
\draw[->] (block_6) edge  (dec_2);
\draw[->] (dec_2) edge ["asymmetric"](block_7);
\draw[->] (block_7) |- (block_6a);
\draw[->] (dec_2) edge ["symmetric"](block_8);
\end{tikzpicture} 
\captionof{figure}{Flow chart of the BRUKER alignment procedure.}
    \label{fig:bruker_align}
\end{minipage}
\begin{minipage}{0.48\textwidth}
\begin{tikzpicture}[CNTRL,
    node distance = 5mm and 15mm,
      start chain = A going below, every node/.style={scale=0.8}
                        ]
\node (block_1) [block, on chain=A]{\begin{tabular}{@{}cc}
     &Align the reference laser\\
     &using input degrees of freedom.
\end{tabular}};
\node (ed_1) [edot, left = of block_1];
\node (dec_1) [cloud, below = of block_1]{\begin{tabular}{@{}cc} & Amplitude A and B\\& above threshold\end{tabular}};
\node (block_2) [block, below = of ed_1]{Align $M_1$};
\node (block_3) [block,below = of dec_1]{Input reference laser};
\node (block_4) [block,below = of block_3]{\begin{tabular}{@{}cc} &Move $M_2$ back and forth\\& and measure SD \& A \& B\end{tabular}};
\node (dec_2) [cloud,below = of block_4]{Phase of SD=A or -B};
\node (block_5) [block,below = of dec_2]{Input continuous light source};
\node (block_6a) [block,below = of block_5]{Measure interferogram};
\node (block_6) [block,below = of block_6a]{Analyze ILS};
\node (dec_3) [cloud, below = of block_6]{ILS};
\node (block_7) [block, left = of dec_2]{Mirror alignment};
\node (block_9) [block, left = of dec_3]{Mirror alignment};
\node (block_8) [block, below = of dec_3]{FTS aligned};
\draw[->] (block_1) edge  (dec_1);
\draw[->] (dec_1) -|  node[pos=0.3, above]{No}(block_2);
\draw[-] (block_2) edge (ed_1);
\draw[-] (ed_1) edge (block_1);
\draw[->] (dec_1) edge  ["Yes"](block_3);
\draw[->] (block_3) edge  (block_4);
\draw[->] (block_4) edge  (dec_2);
\draw[->] (dec_2) edge  ["True"](block_5);
\draw[->] (block_5) edge  (block_6a);
\draw[->] (block_6a) edge  (block_6);
\draw[->] (block_6) edge  (dec_3);
\draw[->] (dec_2) edge ["False"](block_7);
\draw[->] (block_7) |- (block_4);
\draw[->] (block_9) |- (block_6);
\draw[->] (dec_3) edge ["asymmetric"](block_9);
\draw[->] (dec_3) edge ["symmetric"](block_8);
\end{tikzpicture} 
\captionof{figure}{Flow chart of the new alignment procedure.}
    \label{fig:new_align}
\end{minipage}

The procedure can be optimized by taking advantage of the long coherence length of the reference laser. Therefore a part of the light is coupled into a single-mode fiber and fed to the optical axis of the FTS. The optical path can now be aligned while the mirror $M_2$ moves back and forth. The sinusodial signal observed with the science detector (SD) must be identical to the signal at the photodiodes A and B (with a small phase offset due to the electronic signal processing).  Afterwards the steps from the BRUKER alignment procedures using a continuous light source can be used. However, in our experience the ILS is at that point already very symmetric and the phase spectrum very smooth and flat. The signals measured with the SD and photodiodes A and B carry information about further artifacts, e.g. a varying speed of the moving mirror, or a distortion of the rails on which the mirror moves.

The new alignment methods circumvents the observation of the Haydinger fringes and uses the phase coherence of the reference laser. The alignment is thus faster and often more precise than the former method.

\bibliography{00Astro} 
\bibliographystyle{plainnat} 

\end{document}